# Branch Prediction in Hardcaml for a RISC-V 32im CPU

Alex Saveau
asaveau@cs.washington.edu

*Abstract—* Accurate branch prediction is a critical part of high performance instruction stream processing. In this paper, I present a hardware implementation of branch prediction for a RV32IM CPU, starting with static decode stage predictions and culminating in the use of BATAGE. In addition, I detail my experience writing the RTL in Hardcaml, a hardware description library for the functional programming language OCaml.

## I. Introduction

Branch prediction is a crucial part of any high performance processor and continues to rise in importance as instruction windows become larger. In fact, there is a cap on the performance of a processor that does not execute both branch directions simultaneously tied to branch prediction accuracy [1]. Consider a mispredicted branch: it effectively acts as a synchronization barrier, requiring all of its dependencies to be completed before the processor can move forward. By Amdahl's law, if all of those depended-on instructions were completed instantly, the next batch of instructions following the branch would be blocked on the branch instruction's flow through the processor's pipeline, thus limiting performance to the processing time of mispredicted branches (assuming all other bottlenecks have been obviated). Minimizing branch mispredictions is therefore a key method for improving processor performance.

While the concept of branch prediction is simple, an implementation that takes advantage of all possible control transfer related optimizations requires a myriad of techniques in different pipeline stages. Superscalar architectures or strict timing requirements further exacerbate the complexity of accurate branch prediction, for example introducing new opportunities such as predicated instruction inference for short branches [2] or requiring pipelined predictions and predictor updates [3]. This paper's implementation primarily focuses on branch prediction accuracy, but does include enough optimizations to run at 50MHz on an Arty A7-100T FPGA.

To write hardware, the industry primarily uses Verilog or VHDL [4]. While the author is not familiar with VHDL, Hardcaml solves a number of problems with Verilog and introduces new capabilities that are difficult to express in an imperative fashion.

### A. Paper overview

The paper is organized as follows. First, I very briefly introduce the processor, followed by an explanation of each implemented branch prediction technique. Then, I present an evaluation of each technique, followed by future work to further improve the processor. Finally, I describe the key advantages offered by Hardcaml over the course of the processor's development.

## II. CPU overview

The branch prediction described in this paper was incorporated into a RV32IM processor implemented in Hardcaml. The in-order, scalar CPU consists of 4 pipeline stages—the decode and register load stages are combined due to RISC-V instructions always placing rs1 and rs2 addresses in the same bits of any non-compressed instruction. Both instruction and data memory are stored in SRAM with single-cycle access latency, but no other memories are available and therefore no caches are present. A memory mapped UART port is available for external I/O.

Branch prediction verification is performed in the execute stage (thus, instructions in the writeback stage can be considered retired for the purposes of branch prediction) and has a one cycle recovery penalty.

## III. Branch prediction techniques

Different branch prediction techniques must be used depending on the pipeline stage in which the prediction is occuring, whether or not the instruction stream has been encountered before, and the type of control transfer instruction encountered (e.g. a branch vs. a jump). Overall, the goal of a processor's frontend is to keep the backend full with useful instructions.

### A. Static branch prediction

When an instruction stream has never been encountered before, an informed prediction can be made at the earliest in the decode stage. After decoding an instruction, the processor knows what is about to be executed. For branch prediction purposes in RISC-V, there are three instruction types which can be acted upon:

- *Direct unconditional jumps.* A 100% accurate prediction can be made for unconditional jump instructions in the decode stage as the processor has complete knowledge of both the current program counter and jump offset (since it is encoded in the instruction's immediate field).

    Only the tiniest of processors that cannot afford an extra adder should avoid this optimization.

- *Indirect unconditional jumps.* Unfortunately, without additional information, no prediction can be made as the jump is relative to an arbitrary register's unknown value. Checking for the zero register is unlikely to be of use as this means the program is jumping to a target address within the $0 \pm 2^{11}$ range which occurs rarely (if ever) in real systems.

    Processors should stall prior stages until the jump is resolved to avoid wasting power executing known incorrect instructions.

- *Branches.* Branch instructions are in between direct and indirect jumps in terms of predictability. On the one hand, the jump offset is fully known (due to also being encoded in the instruction's immediate field), but on the other, the processor does not know whether or not the branch will be taken.

    Modern compilers, especially when profile guided, can do a good job of statically predicting branch outcomes [5]. Compilers then encode this knowledge in the structure of the program they emit by attempting to maximize repeated and then straight-line execution. Repeated execution ensures the instruction cache remains hot, while straight-line execution makes the next cache line easily prefetchable. As far as branches are concerned, maximizing repeated and straight-line execution implies that if a branch is to be taken, it should point backwards.

    Hardware can therefore predict that branches pointing backwards are likely to be taken, leaving the prediction for other branches as not taken.

### B. Return address stack (RAS)

Returning from a function unfortunately requires indirect jumps as the caller is unknown (if a direct jump could be used, then the function has only one caller and should therefore be inlined). Fortunately, the return address is known dynamically by the fact that the function was called. Thus, hardware can use ISA specified [6] calling conventions to determine when a jump is a call vs. a return (or neither).

On a call, the address of the instruction following the call is pushed onto the RAS. On a return, the top of the RAS is consulted to determine the next fetch program counter.

1) *When to update the RAS:*

Because it is not uncommon to see function calls or returns quickly follow each other (consider any recursive function), updating the RAS at retirement introduces too much delay between the time the prediction is used and consumed. Thus, the RAS should be updated after decoding a call or return.

To handle incorrectly speculated updates, a few options are available [7], two of which are considered depending on the size of the processor's instruction window:

- For deeply pipelined processors that are likely to have many function calls in flight, two stacks will need to be instantiated. The correct instance of the stack will be updated during in-order retirement while the possibly incorrect stack will be maintained in the decode stage. On a misprediction, the entire correct stack is copied from retirement to decode. Depending on the processor's target recovery latency, the stack will likely need to be implemented in flip flops to support fast copies.
- For smaller processors that are unlikely to have multiple function calls in flight, it is enough for a call or return to pass the previous top-of-stack pointer down to retirement. Misprediction recovery then only needs to restore this pointer.

    To understand why this simple approach is so powerful, consider which sequences of pushes and pops irreparably destroy data on the stack: only a pop followed by a push will lose data, specifically the entry that was popped. Pushes (assuming they do not overflow the stack), pops, and pushes followed by pops either do not modify the stack or add data that will be ignored after recovery. Thus, tearing—where the stack has useful entries, but the top-of-stack pointer is off-by-N—is prevented and minimal data loss occurs.

2) *What to do if the stack overflows:*

Because function call trees tend to grow wider as they go deeper (consider the typical pyramid shaped flame graph as opposed to a fence/pillar shaped flamegraph), the total number of function calls will be greater in the deeper part of the call tree. The RAS will therefore be most useful deep in the call stack and, as a consequence, it is preferable to lose older return addresses than it is to prevent new return addresses from being added to the stack.

Hardware should therefore do nothing and let the stack overflow naturally.

### C. Branch target buffer (BTB)

To make predictions as early as possible, the ability to redirect the instruction stream in the fetch stage is necessary. Unfortunately, the fetch stage has no knowledge of the instructions being fetched aside from their addresses. Thus, a memory is needed to keep track of known control transfer instructions. The BTB must store the following information:

- Is the instruction a branch? If so, consult a dynamic predictor.
- Is the instruction a return? If so, consult the RAS.
- Is the instruction a jump? If so, redirect the instruction stream unconditionally.
- What is the target address? That is, where should the instruction stream be redirected to?

Like any cache, there are a number of different implementation strategies [8] with the primary constraint being that the BTB needs to have a latency shorter than the time it takes an instruction to arrive at the decode stage, ideally one cycle. Because RISC-V jumps and branches have different immediate encodings, the BTB can be partitioned such that jumps are stored in one memory and branches are stored in another, thus allowing the branch BTB to only store a 12-bit offset instead of the full program counter needed for jumps.

*1) When to add entries:*

For a given instruction, its contents in the BTB will never change because no dynamic information is stored. As a consequence, updates are not necessary. However, one must choose when new entries should be added to the BTB. Since static decode stage predictions will occur on new instruction streams, adding entries before retirement only serves to eliminate wasted cycles between the fetch and decode stages when decoding a tight loop, but provides no benefit otherwise as the true outcome of the branch is unknown. Furthermore, adding an extra write port is likely to be a poor tradeoff as this increases the complexity of the BTB which ideally needs to maintain single-cycle access latency. Experimentally, I observed slight improvements in hand-written assembly programs when updating the BTB after decode, but saw regressions in programs generated by LLVM.

Hardware can therefore stick to retirement-only BTB insertions.

There is one final consideration: should non-taken branches be added to the BTB? If a branch is never taken, it is clearly unhelpful to store the branch in the BTB as the fetch stage will naturally continue straight-line execution. However, the decode stage may believe the branch will be taken and cannot know that the branch has been seen before to be not taken without an entry in the BTB. Experimentally, adding mispredicted, non-taken branches to the BTB decreases performance due to evicting useful BTB entries, which matches the results found by Bray et al. [9]. The Intel Pentium processor is also documented as only adding taken branches to the BTB [10].

Thus, only taken branches should be added to the BTB to augment the decode stage's knowledge of the instruction stream.

*D. Dynamic predictors*

A branch target buffer provides little benefit without a dynamic predictor, as stored branches are assumed to be taken, thus overriding static predictions. For example, a branch that is statically predicted as not taken, but dynamically taken some small number of times will be added to the BTB and forever predicted as taken even though this decreases accuracy. One could remove entries from the BTB on a misprediction, but that would be equivalent to implementing a last action predictor.

Instead, a dynamic predictor which learns branch behavior should be consulted on a BTB hit, thereby overriding static predictions only after having seen the branch before and thus being able to make a more informed prediction.

This paper's processor implements BATAGE [11] which will only briefly be described here. The key ideas are as follows:

- Local and global branch histories are correlated to improve accuracy.
    - Local history refers to the past behavior of one specific branch.
    - Global history refers to the past behavior of all branches.
- A list of bimodal-like entries is retrieved to make a prediction, indexed by hashing a combination of the program counter and some geometrically increasing subset of the global history vector.
    - Bimodal predictors are a generalization of last action predictors with a delay between the transition from one prediction to another. That is, they can be thought of as holding a certain mass that must be moved across the taken/not taken prediction boundary, requiring effort proportional to the distance of the mass from the boundary to change its prediction.
- Each entry is given a confidence score and the highest confidence entry is used to make the prediction.
    - To break ties, entries indexed by a larger subset of the global history are given priority.

*1) Concrete example:*

Consider Program 1, which increments a loop counter and tracks the number of times it was even vs. odd.

A TAGE-like predictor is able to achieve 100% accuracy on this program whereas bimodals will consistently mispredict. Notice the global history pattern of branches (including jumps) split up by loop iteration:

```
011 111 011 111 011 111 ...
```

Execution 1: Flip flopping branch

By partitioning our prediction on this global history, we can predict taken after having seen 011 and not taken after having seen 111. This ability to partition predictions on the global branch history is TAGE's superpower.

```
    li a0, 50           ; Loop counter

loop:
    andi a1, a0, 1
    bne a1, zero, odd ; Studied branch
    j even

odd:
    addi t1, t1, 1
    j check

even:
    addi t0, t0, 1

check:
    addi a0, a0, -1
    bnez a0, loop
```

Program 1: Perfect predictability with TAGE

*2) When to update predictor structures:*

Predictors must be trained with correct information which can only be obtained during retirement, thus updates must occur during in-order retirement.

However, the global history used when making a prediction must match the history used for retirement updates to be useful. Thus, the global history must be updated speculatively, but thankfully, implementing the history as a circular buffer and restoring the write pointer on a misprediction makes recovery trivial. Note that for a target maximum history length $H$, the circular buffer must be larger than $H$ by at least the number of control transfer instructions expected to be in flight or, conservatively, the entire instruction window size. These extra entries are necessary to avoid losing the oldest history entries to speculative writes.

*3) Pipelining:*

Keeping the backend full requires a continuous, useful instruction stream, which can be accomplished by making predictions of increasing accuracy for a given instruction as it flows through a processor's pipeline stages. So long as the earlier pipeline stages are over-provisioned, correcting an earlier prediction should have minimal impact on the amount of work available to the backend. Therefore, processors with strict timing requirements such as Berkley's SonicBOOM [2] employ tiered predictors: a fast predictor is first used in the fetch stage followed by the arrival of a slower but more accurate pipelined prediction (such as TAGE) in the decode stage, used to correct the earlier fetch stage prediction if necessary.

Because this paper's processor has just 4 pipeline stages and only a single instruction is fetched at a time, delaying predictions by one cycle was deemed too slow and I therefore chose to use the BATAGE predictor in the fetch stage. To further improve accuracy, I also use BATAGE's prediction in the decode stage if the BTB missed and the prediction came from one of BATAGE's tagged banks, as they effectively act as a BTB entry (the branch must have been seen before if it was allocated an entry in a tagged bank).

## IV. EVALUATION

Each technique described above has been implemented in this paper's processor and evaluated on a number of sample programs written in Rust, compiled to rv32im assembly with LLVM, and linked by LLD using the `relax` option to convert function call indirect jumps to direct jumps. Note that each implemented technique builds upon the others. Sample program dynamic instruction counts range from the low thousands to hundreds of millions for the donut program. A brief description of each program is provided in Table 1.

Figure 1 evaluates the overall performance improvements achieved by each branch prediction technique as measured by IPC (i.e. instructions retired divided by clock cycles). Figure 2 and Figure 3 evaluate the misprediction rate per 1000 instructions for jumps and branches respectively. Finally, Figure 4 offers a different perspective on the branch misprediction rate by instead reporting branch prediction accuracy.

|  | **Description** |
|---|---|
| **Calculator** | A program which accepts two operands and an operation as strings, parses them, and outputs the result. |
| **Donut** | The famous ASCII donut rendering program [12]. |
| **Fibonacci** | A recursive implementation of fibonacci which outputs the nth number in the sequence as specified by a number parsed from a string. |
| **Pi** | A program which computes $\pi$ using the Bailey–Borwein–Plouffe formula. |
| **Rng** | A program which emits N random numbers using the Xoroshiro256++ algorithm. |
| **Sort** | A program which accepts N strings, parses them as signed numbers, sorts them, and then emits them back. |
| **Word count** | A program which, given input text, emits the number of characters, words, and paragraphs present in the input. |

Table 1: Sample program descriptions

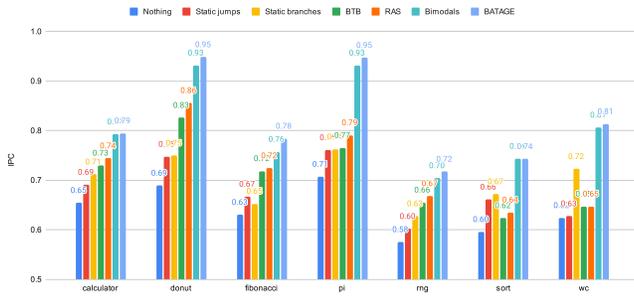

Figure 1: Overall performance comparison

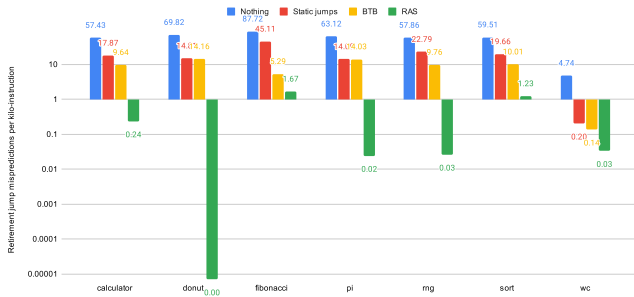

Figure 2: Jump misprediction rate comparison

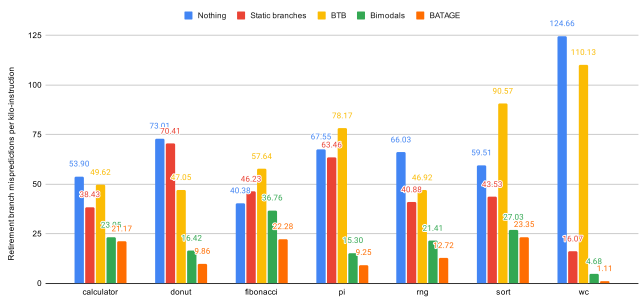

Figure 3: Branch misprediction rate comparison

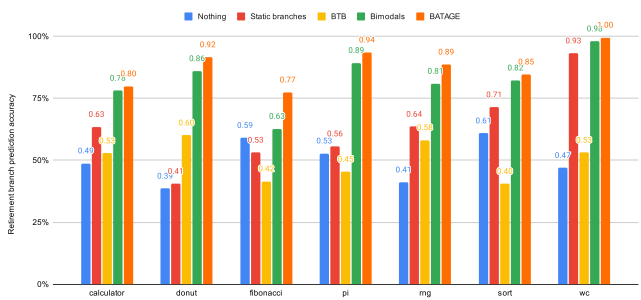

Figure 4: Branch prediction accuracy comparison

As expected, static jump predictions always improve performance. Static branch predictions often, but don't always improve performance—with profile guided optimizations (PGO), I believe the compiler could significantly improve its predictions, or at the very least, not worsen performance. Thus, I believe almost all processors should implement static predictions as the hardware cost of doing so is basically free and the performance gains are notable.

Figure 2 demonstrates the importance of including a RAS: once added, jump mispredictions are all but eliminated. Due to being ruthlessly effective, I also believe most processors should include a RAS (unless the expected programs make minimal use of function calls).

The BTB improves performance of jumps, especially in function call heavy programs such as *Fibonacci*, but as noted earlier, acts as an always taken predictor for branches, thereby worsening performance. Once a dynamic predictor is introduced (i.e. bimodals), all lost performance is recovered and then further improved.

Finally, BATAGE shows significant accuracy gains, but limited overall performance improvements due to approaching the limits of available performance improvement from more accurate branch prediction (see Figure 5). The author would like to note that this paper's implementation of BATAGE does not match the expected ~4 mispredictions per kilo-instruction claimed in the paper. Further investigation is needed to determine the cause: bugs in the implementation are a likely candidate, but it may be possible that this paper's evaluation programs are less predictable than those used to benchmark BATAGE.

### A. Future work

As Figure 5 shows, there are still (small) improvements left on the table. This section proposes some ideas to address current shortcomings.

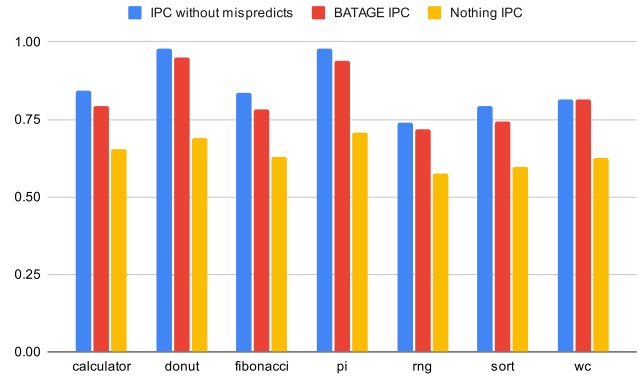

Figure 5: IPC if all cycles lost to mispredictions were avoided

#### 1) *Pipelining:*

As a consequence of using BATAGE predictions in the fetch stage, the number of memory banks used is limited to what can be processed in a single cycle. It would be preferable to overprovision the fetch stage such that it outputs multiple instructions per cycle and pipeline the BATAGE prediction for arrival in the decode stage, thereby allowing larger TAGE structures without leaving the decoder empty on a corrected prediction.

2) *More varied sample programs:*

The sample programs currently used for benchmarking purposes are limited and should be expanded to encompass more application domains.

I also did not test any programs compiled with PGO enabled, but confirming the performance advantage of dynamic predictors over even perfect static predictions would be interesting.

3) *Sample program prediction accuracy model checking:*

To determine if this implementation of BATAGE is performing as expected, execution traces for the evaluation programs must be produced and fed into the official BATAGE model to produce expected misprediction rates. This paper's evaluation can then be put in context with the BATAGE paper's results.

4) *Static indirect unconditional jumps:*

This paper's implementation does not stall the frontend when an indirect jump is encountered for which we have no information, but stalling would save power and avoid a one cycle misprediction recovery latency.

Additionally, a variant of TAGE can be used to predict indirect jumps [13] which should be incorporated if the processor is expected to handle dynamic languages or programs with virtual function calls.

## V. Hardcaml

> Reader beware: this section is primarily opinion and experience based.

The industry primarily uses Verilog or VHDL as the hardware description language for digital design, with Verilog being the more popular choice.

### A. Verilog shortcomings

I have (limited) experience with Verilog from university digital design courses, but can attribute my aversion to the language to three broad categories: predictability, safety, and expressiveness.

1) *Predictability:*

In this context, predictability refers to the presence of implicit behavior. When a tool is unpredictable, its users have to fight their way to a given goal—it will feel like the tool is in control of the process rather than the other way around. Verilog exhibits several behaviors that make it feel unpredictable:

- The designer's wish for a register, wire, or latch is inferred, leading to unexpected behavior when the designer makes a mistake that causes Verilog's inferred construct to differ from that which was desired. Adding salt to the wound, Verilog offers a `reg` keyword which does not, in fact, create a register and instead continues to rely on the presence of an assignment in either a combinational or sequential block to determine if the signal should be a wire or a register.
- Verilog supports imperative loops, which, while potentially convenient for a designer, do not always map well into hardware. For example, it is easy in Verilog to write loops with dependencies or that terminate early, thereby introducing unexpectedly large gate delays.
- Memory inference in Verilog is tricky. A designer who wishes to remain platform agnostic cannot explicitly create an SRAM and must instead rely on the tools to correctly infer an SRAM based on access patterns to a large array.

In general, Verilog code is not guaranteed to be synthesizable which is upsetting considering Verilog is supposed to be a hardware description language.

2) *Safety:*

Safety here means that the tools are working for you to help prevent mistakes, rather than assuming the programmer is always right. The persistent prevalence of buffer overflow vulnerabilities in C/C++ code has largely disproven the theory that humans are capable of writing 100% correct code when no safeguards are present. Furthermore, tools that inform you of a mistake are a good thing: we should be spending our time solving hard, interesting problems, not debugging trivially detectable bugs by well designed tools. Verilog has a number of safety issues:

- Wires may be unassigned, leading to unknown hardware behavior.
- Port widths are not checked, leading to arbitrary resizing, or worse, unexpected truncation.
- Combinational loops are allowed by default, which is almost never the designer's intention.
- Test bench simulations require the designer to correctly simulate a fake notion of time which is frustrating boilerplate to write at best, and leads to incorrect simulations that do not match hardware at worst.

In general, Verilog follows the old style of opt-in safety rather than the modern opt-out safety.

3) *Expressiveness:*

While a tool may be theoretically capable of accomplishing any task, some tasks may be easier to describe than others. Expressiveness here refers to the ability to write clear and concise RTL.

Verilog has some arcane syntax that can be off-putting, but primarily suffers from its imperative coding paradigms. Hardware is fundamentally functional in nature which means expressing combinational transformations is much easier when written functionally. This does not mean hardware should never be described imperatively, but rather that there should always be an option to describe hardware functionally, which is the ability Verilog lacks.

*B. Why Hardcaml instead of other HDLs?*

When I set out to find a new HDL, I considered a few other languages of note:

- HDLs whose source language is C. C as a programming language lacks both safety and expressiveness and is therefore not considered an improvement.
- Python and Haskell based HDLs. These HDLs attempt to make writing hardware just like writing software, thereby straying even further from predictability than Verilog. I am skeptical that efficient hardware can be generated without control over the netlist.
- Chisel. Chisel does not validate port widths which may seem like a small gripe, but I encountered several dozen port width mismatch errors over the course of development that likely would have resulted in hours of wasted debugging time had they not been caught. That said, of all the HDLs out there, Chisel has the most potential for displacing Verilog and VHDL—Chisel is backed by Berkley and the CHIPS Alliance. On top of that, Google has already [built an Edge TPU with Chisel](#).

*C. Why Hardcaml?*

Hardcaml largely solves all of the aforementioned issues. The simplicity of the core library is striking: one can instantiate primitive operations, wires, registers, and muxes. That's it.

While this simplicity is initially frightening, it quickly becomes liberating when composed with other features, giving the designer absolute control over the generated netlist. Because Hardcaml is a library for Ocaml, all features are implemented with normal Ocaml constructs such as function calls. Register and wires are therefore simply functions that return Hardcaml's `Signal` data type. Muxes and primitives accept signals and return new signals. This means that producing combinational logic is as simple as writing a function.

Consider the real piece of code in RTL 1 used to generate 5 PRNGs that match Procedure 1 from the BATAGE model. The code is both clear and concise:

1. We first define a combinational logic function `gen_random` which receives a signal `x` and randomizes it via XORs and shifts before returning it.
2. Next we define a function `rng` which produces a register with sequential logic that computes `gen_random` every cycle, parameterized by a starting value.
3. Finally, we instantiate 5 different PRNGs, each with different starting values.

In Verilog, I would not be confident that no mistakes had been made, therefore I probably would have written a test bench for this module. In Hardcaml, I do not know how you could make a mistake in this RTL.

```c
uint32_t rando() {
    // Marsaglia's xorshift
    static uint32_t x = 2463534242;
    x ^= x << 13;
    x ^= x >> 17;
    x ^= x << 5;
    return x;
}
```

Procedure 1: The BATAGE model's PRNG

```ocaml
let randoms ~clock ~clear ~enable =
  let open Signal in
  let gen_random x =
    (* sll stands for shift left logical *)
    let x = x ^: sll x 13 in
    let x = x ^: srl x 17 in
    let x = x ^: sll x 5 in
    x
  in
  let rng ~start =
    let width = 32 in
    (* Create a register whose next value is
       produced by the function gen_random
       and is cleared to the constant start. *)
    reg_fb
      ~enable
      ~width
      ~f:gen_random
      (Reg_spec.create ~clock ~clear ()
       ▷ Reg_spec.override
           ~clear_to:(of_int ~width start))
  in
  (* Create 5 registers, each with its own seed. *)
  ( rng ~start:2463534242
  , rng ~start:1850600128
  , rng ~start:3837179466
  , rng ~start:4290344314
  , rng ~start:0614373416 )
;;
```

RTL 1: A function which returns 5 PRNG signals

*D. Model checking*

There are a number of other examples that showcase Hardcaml's power, but a full paper could be written on the subject so I will only further discuss a key benefit Hardcaml provided while implementing BATAGE: interactive model checking.

The combinational logic portions of this paper's BATAGE implementation are model checked: the BATAGE model was reimplemented in Ocaml and used to validate the correctness of this paper's BATAGE implementation by generating random inputs to the circuit and validating them against the model. Dozens and dozens of bugs were found through this process! While the same thing could have presumably been achieved using Verilator, having the model and implementation in the same language made it trivial to add print statements and use waveform debugging to quickly identify the source of the bugs.

## VI. CONCLUSION

In this paper, I presented advanced branch prediction techniques implemented for a RV32IM processor, written in Hardcaml. Branch prediction continues to play a key role in enabling high performance instruction stream processing as evidenced by the IPC gains achieved with BATAGE. Future work remains to improve the processor (and its branch prediction), but there is now a solid foundation to build upon.